\documentclass[english]{article}
\usepackage{lmodern}
\usepackage[T1]{fontenc}
\usepackage[latin9]{inputenc}
\usepackage{amssymb}
\usepackage{graphicx}
\usepackage{esint}
\usepackage{babel}
\makeatletter
\newcommand{\lyxaddress}[1]{
\par {\centering #1
\vspace{1.4em}
\noindent\par}
}

\newcommand{\lyxemail}[1]{
\par {\centering #1
\vspace{1.4em}
\noindent\par}
}
\makeatother

\usepackage{babel}
\begin{document}

\title{Emergence of spaces and the dynamic equations of FRW universes in
the $f(R)$ theory and deformed $\textrm{Ho}\check{\textrm{r}}\textrm{ava}$-Lifshitz
theory}
\author{Fei-Quan Tu, Yi-Xin Chen\footnote{Corresponding author}}
\date{}
\maketitle

\lyxaddress{Zhejiang Institute of Modern Physics, Department of
Physics, \\ Zhejiang University, Hangzhou, 310027, P. R. China}
\lyxemail{Email: fqtuzju@foxmail.com, yxchen@zimp.zju.edu.cn}

\begin{abstract}
It has been shown that Friedmann equation of FRW universe can be derived
from the idea which says cosmic space is emergent as cosmic time progresses
and our universe is expanding towards the state with the holographic
equipartition by Padmanabhan. In this note, we give a general relationship
between the horizon entropy and the number of the degrees of freedom
on the surface, which can be applied to quantum gravity.
We also obtain the corresponding dynamic equations
by using the idea of emergence of spaces in the $f(R)$ theory and
deformed $\textrm{Ho}\check{\textrm{r}}\textrm{ava}$-Lifshitz(HL)
theory.
\end{abstract}

\section*{1. Introduction}

Since the discovery of black hole thermodynamics\cite{key-1,key-2}in
the 1970s, physicists have accepted the relationship between thermodynamics
with horizon and gravity theory. In 1995, Jacobson\cite{key-3}argued
that the Einstein equation is derived from the proportionality of
entropy and horizon area together with the Clausius relation $\delta Q=TdS$
with $\delta Q$ and $T$ interpreted as the energy flux and Unruh
temperature seen by an accelerated observer just inside the horizon,
and pointed out that the Einstein equation is an equation of state.
Padmanabhan\cite{key-4}notices the Einstein's equation in the spherically
symmetric spacetime can have a similar form as the first law of thermodynamics
of the black hole horizon.

Verlinde\cite{key-5}points out that gravity is explained as an entropic
force caused by changes in the information associated with the positions
of material bodies. With the assumption of the entropy force, the
local Unruh temperature, the holographic principle, and the equipartition
law of energy, combining with the definition of the Komar mass, he
obtains the Einstein equation, where the Komar mass plays an important
role.

In the FRW model, it has been shown that the Friedmann equation which
describes the dynamic of the isotropic and homogeneous universe can
be derived from the first law of thermodynamics with the apparent
horizon, by assuming a temperature $T=\frac{1}{2\pi r_{A}}$ and the
entropy $S=\frac{A}{4}$, where $r_{A}$ and $A$ are radius and area
of the apparent horizon. Cai and Kim obtain the Friedmann equation
by using entropy formula for the static spherically symmetric black
hole horizon in Gauss-Bonnet gravity and in more general Lovelock
gravity\cite{key-6}. The similar results are also obtained in scalar-tensor
gravity and $f(R)$ gravity\cite{key-7}, and deformed $\textrm{Ho}\check{\textrm{r}}\textrm{ava}$-Lifshitz(HL)
gravity\cite{key-8}.

It is generally accepted that gravity is described as an emergent phenomenon like
fluid mechanics or elasticity in recent years. Padmanabhan\cite{key-9}suggested
an idea that cosmic space is emergent as cosmic time progresses. Further
he argued that the difference between the number of the surface degrees
of freedom and that of the bulk degrees of freedom in a region of
space drives the accelerated expansion of the universe through a simple
equation $\Delta V=\Delta t(N_{sur}-N_{bulk})$ , where $V$ is the
Hubble volume in Planck units and $t$ is the cosmic time in Planck
units, derived the standard Friedmann equation of FRW universe. The
Gauss-Bonnet gravity and more general Lovelock gravity are also studied\cite{key-10}.
And emergent perspective of gravity is further investigated\cite{key-11}.

In this paper, we shall propose a general relationship about degrees of freedom on the surface, which is described by the relation between the number of degrees of freedom on the surface and the entropy of horizon. The relation given by Padmannabhan is based on the thought that degrees of freedom on the surface uniformly distribute over the area. But the thought isn't valid in some theories such as quantum gravity. And his relation can be reduced by our relation in the GR theory. Therefore our relation is important and general to investigate the expansion of the universe in general gravity theories. Then we use the
$f(R)$ theory and deformed HL gravity as examples to explain why
the relation given by Padmanabhan is invalid but our relation is valid
and obtain the corresponding dynamic equation of FRW universe.

This paper is organized as follows. In Section 2 we review the relationship
between the thermodynamics of the black hole horizon and spacetime
horizon , then we give a relation about the surface number of the
degrees of freedom. In Section 3 we simply recall the Padmanabhan's
idea about the emergence, and show the dynamic equation of FRW universe
in the $f(R)$ theory and deformed HL gravity theory under our relation.
Section 4 is for conclusions and discussions. The units are chosen with
$c=\hbar=1$.

\section*{2. Description of degrees of freedom on the surface}

The equivalence principle tells us that effects of gravitational field
is locally indistinguishable from the effects of accelerated motion
in the flat spacetime. In fact, all notions of thermodynamics are
observer dependent. A local inertial observer (local falling freely
observer)will not attribute to spacetime a temperature, and a local
Rindler observer(local accelerated observer)with respect to local
inertial observer will attribute to spacetime a non-zero temperature.
Padmanabhan has found these Rindler observers around any event will
attribute not only a temperature but also an entropy to the horizon
in a static, spherically symmetric spacetime, and the first law of
thermodynamics is also satisfied\cite{key-12}. In the black hole
physics, an observer who freely falls through the horizon will not
attribute a temperature to the black hole while the observer who is
stationary outside the black hole horizon will attribute a temperature
to the black hole. Hence the relationship between the observer at
rest outside the black hole horizon and the freely falling observer
is exactly the same as the relationship between an Rindler observer
and an inertial observer in flat spacetime\cite{key-11}.

Now we consider the flat spacetime for a class of Rindler observers,
the metric is $ds^{2}=-dT^{2}+dX^{2}+dL_{\perp}^{2}$. Using transformation
of coordinates, the metric can be expressed in the form $ds^{2}=-2\kappa ldt^{2}+\frac{dl^{2}}{2\kappa l}+dL_{\perp}^{2}$,
where $\kappa$ is an acceleration. The form is called the Rindler
metric. The Rindler metric is similar as that of black hole in
the spherically symmetric spacetime if we let $2\kappa l\equiv f(r)$
(for the review\cite{key-13}).

As we know, the number of degrees of freedom are described by the
entropy in the statistical physics, and an accelerated observer attributes
an entropy to the horizon of spacetime. Hence we can use
\begin{equation}
N_{sur}=\frac{4S_{H}}{k_{B}}
\end{equation}
to describe the number of the degrees of freedom on the surface of
the horizon, where $S_{H}$ is the entropy of the horizon. Because
the relation between the entropy $S_{H}$ and the area of its horizon
is $S_{H}=\frac{k_{B}A}{4L_{P}^{2}}$ in the GR theory and some other
theories, Eq.(1) can be turned into
\begin{equation}
N_{sur}=\frac{A}{L_{p}^{2}}
\end{equation}
 in those theories, where $L_{P}=\sqrt{\frac{\hbar G}{c^{3}}}$ is
the Planck length. But in some gravity theories like the quantum gravity
with $S_{H}\neq\frac{k_{B}A}{4L_{P}^{2}}$, Eq.(2) isn't valid.

An observer very close to the event horizon, performing local experiments
at length scales small compared to curvature scale, has no way of
distinguishing between a Rindler coordinate system in a flat spacetime
and the black hole spacetime\cite{key-12}. We can reduce the entropy
of a Schwarzschild's black hole to one of accelerated observers in
the flat spacetime by taking the limit of the black hole of very large
mass\cite{key-11}. Therefore the entropy of horizon $S_{H}$ can
be formally replaced by that of a black hole $S$. So the relation
between the number of degrees of freedom on the surface and the entropy
form of a black hole is
\begin{equation}
N_{sur}=\frac{4S}{k_{B}}.
\end{equation}

\section*{3. Emergence of space in the $f(R)$ theory and deformed HL theory }

First, let us recall Padmanabhan's work\cite{key-9}. He thinks
that cosmic space is emergent as cosmic time progresses. To understand
it, he uses a specific version of the holographic principle. For a
pure de Sitter universe with a Hubble constant $H$, the holographic
principle is expressed in the form (called holographic equipartition)

\begin{equation}
N_{sur}=N_{bulk},
\end{equation}
where the $N_{sur}$ is the number of degrees of freedom on the spherical
surface given by Eq.(2) and $A=\frac{4\pi}{H^{2}}$ is the area with
Hubble radius $H^{-1}$. Here we would like to make a note. Hubble
constant $H$ is a fundamental constant in the cosmology, and its
dimension is reciprocal of the length, hence we use $H^{-1}$as the
radius of the expanding universe can be accepted. He takes the effective
number of the bulk degrees of freedom which obey the equipartition
law of energy
\begin{equation}
N_{bulk}=|E|/(1/2)k_{B}T
\end{equation}
and the horizon temperature
\begin{equation}
k_{B}T=H/2\pi.
\end{equation}
He takes $|E|$ to be the Komar energy $|(\rho+3p)|V$ contained
inside the Hubble volume
\begin{equation}
V=4\pi/3H^{3}.
\end{equation}
For the pure de Sitter universe $p=-\rho$, then one can obtain the
standard result
\begin{equation}
H^{2}=8\pi L_{p}^{2}\rho/3.
\end{equation}

He thinks that the expansion of the universe is being driven towards
holographic equipartition(Eq.(4)), the equation is
\begin{equation}
\frac{dV}{dt}=L_{p}^{2}\left(N_{sur}-N_{bulk}\right).
\end{equation}
Using the above definition of $V$, $N_{sur}$, $T$ and $N_{bulk}$
, one obtains the relation
\begin{equation}
\frac{\ddot{a}}{a}=-\frac{4\pi L_{p}^{2}}{3}(\rho+3p).
\end{equation}
This is the standard dynamical equation of  FRW universe filled by
perfect fluid with energy density $\rho$ and pressure $p$. Using
the continuity equation $\dot{\rho}+3H(\rho+p)=0$, one gets the standard
Friedmann equation
\begin{equation}
H^{2}+\frac{k}{a^{2}}=\frac{8\pi L_{P}^{2}}{3}\rho,
\end{equation}
where $k$ is an integration constant, which can be interpreted as
the spatial curvature of  FRW universe. In order to have the asymptotic
holographic equipartition, he takes $(\rho+3p)<0$. This implies the
existence of the dark energy. The existence of the dark energy derives
our universe towards the state with the holographic equipartition.
His idea about the space is emergent as time processes and our universe
is expanding towards the state with the holographic equipartition
is creative. However he uses the idea in which the degrees of freedom on
the surface uniformly distribute over the area to obtain $N_{sur}$(see
Eq.(2)), we think that the relationship isn't general and valid
in some theories, especially in the theory of quantum gravity. Subsequently,
we shall consider some examples of such gravity theories.

In the $f(R)$ theory, we will obtain a nontrivial gravitational energy-momentum
tensor due to the existence of high-order curvature, which will result
in a nontrivial energy density and pressure. In the deformed HL gravity,
it also has a nontrivial gravitational energy-momentum tensor due
to the existence of high-order curvature, the cosmological constant, and
the dark radiation. But the gravitational energy density and pressure
don't appear in the GR theory. Hence it is an interesting and meaningful
topic to discuss the evolution of universe under this gravitational
background.

Now we consider emergence of space in the $f(R)$ theory. We take
$|E|$ to be the Komar energy which is the correct source for
gravitational acceleration\cite{key-14}
\begin{equation}
|E|=2\left|\int_{\Sigma}\left(T_{\mu\nu}-\frac{1}{2}Tg_{\mu\nu}\right)U^{\mu}U^{\nu}dV\right|=\left|\rho+3p\right|V,
\end{equation}
where $T_{\mu\nu}$ is the total energy-momentum tensor which contains
two parts, the energy-momentum tensor of matter $T_{\mu\nu}^{(m)}$
and gravity $T_{\mu\nu}^{(g)}$. One can obtain the total energy
density $\rho$ and total pressure $p$ through $T_{\mu\nu}$. Here
we take the black hole entropy, its form is\cite{key-15}
\begin{equation}
S=\frac{k_{B}Af_{R}(R)}{4G}=\frac{k_{B}Af_{R}(R)}{4L_{P}^{2}}.
\end{equation}
Thus from the Eq.(3) we obtain the number of the surface degrees of
freedom
\begin{equation}
N_{sur}=\frac{Af_{R}}{L_{P}^{2}},
\end{equation} where $f_{R}(R)$ denotes that $f(R)$ derivatives with respect to
$R$.
Using Eq.(5), (6) and (7), we have
\begin{equation}
N_{bulk}=\frac{-16\pi^{2}(\rho+3p)}{3H^{4}}.
\end{equation}
According to the Eq.(9), we obtain the dynamic equation
\begin{equation}
\dot{H}+f_{R}H^{2}=\frac{-4\pi L_{p}^{2}}{3}(\rho+3p).
\end{equation}
It is the formal dynamic equation of FRW universe from emergence of
space in the $f(R)$ theory.

Up to now, we have obtained the formal dynamic equation(16) of FRW
universe by the idea of emergence of space in the $f(R)$ theory.
However, we also know the total energy-momentum tensor $T_{\mu\nu}$
can't be determined if we don't give the specific form of the theory, because it has different energy-momentum tensors in the different
gravity theories.

On a large scale of universe, we know that it is isotropic and
homogenous through astronomical observation. we call it cosmological
principle. Under the principle, we can obtain the FRW metric
\begin{equation}
ds^{2}=-dt^{2}+a^{2}(t)\left(\frac{dr^{2}}{1-kr^{2}}+r^{2}(d\theta^{2}+\sin^{2}\theta d\varphi^{2})\right),
\end{equation}
where $k$ is the spatial curvature constant $k=+1$, $0$ and $-1$
corresponding to a closed, flat, and open universe. Here we point
out that the cosmological principle is based on astronomical observations,
therefore we take the metric(17) reasonable in order to determine
$T_{\mu\nu}$. According to the cosmological principle, we take the
form\cite{key-16}
\begin{equation}
T_{00}=\rho(t),T_{0i}=0,T_{ij}=a^{2}(t)\delta_{ij}p(t),
\end{equation}
where $i,j$ run over $1,2,3.$

Now we will show how to determine the total energy-momentum tensor.
In the $f(R)$ theory, the Einstein-Hilbert(EH) action can be written
as
\begin{equation}
S=\int d^{4}x\sqrt{-g}\left(f(R)+2\kappa^{2}L_{m}\right),
\end{equation}
where $\kappa^{2}=8\pi G$. Using the variational principle $\delta S=0$,
we obtain
\begin{equation}
R_{\mu\nu}f_{R}(R)-\frac{1}{2}g_{\mu\nu}f(R)+g_{\mu\nu}\nabla^{2}f_{R}(R)-\nabla_{\mu}\nabla_{\nu}f_{R}(R)=\kappa T_{\mu\nu}^{(m)},
\end{equation}
where $T_{\mu\nu}^{(m)}$ is the energy-momentum tensor of the
matter. Thus we have
\begin{equation}
G_{\mu\nu}f_{R}(R)=\kappa^{2}T_{\mu\nu}^{(m)}+\frac{f(R)-Rf_{R}(R)}{2}g_{\mu\nu}+\nabla_{\mu}\nabla_{\nu}f_{R}(R)-g_{\mu\nu}\nabla^{2}f_{R}(R),
\end{equation}
where $G_{\mu\nu}=R_{\mu\nu}-\frac{1}{2}g_{\mu\nu}R$ is the Einstein
tensor.

We assume the matter to be the perfect fluid, then the energy-momentum
tensor of the matter is
\begin{equation}
T_{\mu\nu}^{(m)}=(\rho_{m}+p_{m})U_{\mu}U_{\nu}+p_{m}g_{\mu\nu}
\end{equation}
with $g_{\mu\nu}U^{\mu}U^{\nu}=-1$. Furthermore, we define
\begin{equation}
T_{\mu\nu}^{(g)}=\frac{1}{f_{R}(R)}\left[\frac{f(R)-Rf_{R}(R)}{2}g_{\mu\nu}+\nabla_{\mu}\nabla_{\nu}f_{R}(R)-g_{\mu\nu}\nabla^{2}f_{R}(R)\right].
\end{equation}
It stands for the energy-momentum tensor of the gravity caused by
higher order derivative.

Therefore we obtain
\begin{equation}
G_{\mu\nu}=\kappa^{2}\left(\frac{1}{f_{R}(R)}T_{\mu\nu}^{(m)}+\frac{1}{\kappa^{2}}T_{\mu\nu}^{(g)}\right)\equiv\kappa^{2}T_{\mu\nu}.
\end{equation}

According to the form(18), and
\begin{equation}
T_{00}^{(m)}=\rho_{m}(t), T_{00}^{(g)}=\rho_{g}(t),
T_{ij}^{(m)}=a^{2}(t)\delta_{ij}p_{m}(t), T_{ij}^{(g)}=a^{2}(t)\delta_{ij}p_{g}(t),
\end{equation}
we obtain
\begin{equation}
\rho=\frac{1}{f_{R}}\left[\rho_{m}+\frac{1}{\kappa^{2}}\left(\frac{Rf_{R}-f}{2}-3Hf_{RR}\dot{R}\right)\right],
\end{equation}
and
\begin{equation}
p=\frac{1}{f_{R}}\left[p_{m}+\frac{1}{\kappa^{2}}\left(\frac{f-Rf_{R}}{2}+f_{RRR}\dot{R}^{2}+f_{RR}\ddot{R}+2Hf_{RR}\dot{R}\right)\right].
\end{equation}
in FRW universe, where $\dot{R}$ denotes that $f(R)$ derivatives with respect to
$t$. By substituting Eq.(26) and Eq.(27) into Eq.(16),
we obtain the dynamic equation of FRW universe from the idea of emergence of space.
Here we shall give some remarks on this result. In these papers\cite{key-17},
the dynamic equation of FRW universe $\dot{H}+H^{2}=\frac{-4\pi L_{p}^{2}}{3}(\rho+3p)$
is obtained in the $f(R)$ theory through the Einstein's equation.
However, Eq.(9) globally describes the expansion of the universe while
the Einstein\textquoteright{}s equation locally describes it. Therefore,
it will lead to a global correction to the area in the entropy when
we use Eq.(9) to describe emergence of space in the $f(R)$ theory.
That is to say, we obtain the global description of the expansion
of the universe in the $f(R)$ theory. It has $f_{R}=1$, $\rho_{g}=p_{g}=0$
when $f(R)=R$, then we obtain the form
\begin{equation}
\dot{H}+H^{2}=\frac{-4\pi L_{p}^{2}}{3}(\rho_{m}+3p_{m}),
\end{equation}
where $\frac{\ddot{a}}{a}=\dot{H}+H^{2}$. we can see that it is nothing
but the standard dynamic equation in the GR theory.

Our conclusion is presented as follows. Eq.(16) can be reduced to
Eq.(28) when $f(R)=R$ , so we confirm Eq.(16) is the correct result
in the $f(R)$ theory. we can also see that in the paper\cite{key-7},
the corresponding dynamic equation can't be reproduced from the entropy
formula (13) in the $f(R)$ theory , the ansatz for the temperature
on the apparent horizon and the first law of thermodynamics, so we think that our work
can better describe the expansion of universe based on the $f(R)$
gravity. Moreover, we can't use the Eq.(2) to obtain the dynamic equation
of FRW universe because the entropy $S$ isn't proportional to the
area $A$. Here we use the Einstein's equation to determine the total
energy-momentum tensor because it can't be determined if we don't
give the specific form of the theory, but we don't use the specific
form of the Einstein tensor. Therefore it is more convenient to determine
the dynamic equation of the FRW universe. In this process, we should
pay attention to the definition of Komar mass, and have to take the
total energy density and total pressure in the dynamic equation of
FRW universe, because it contains not only the effects of matter,
but also the effects of higher-order gravity.

In the deformed HL gravity, we also take the form of Komar energy
$|E|=\left|\rho+3p\right|V$ as the $f(R)$ theory, the entropy
has the form\cite{key-18}
\begin{equation}
S=\frac{k_{B}A}{4G}+\frac{k_{B}\pi}{\omega}\ln\frac{A}{4G},
\end{equation}
where the parameter $\omega=16\mu^{2}/\kappa^{2}$. Using the above
formula (5), (6) and (7), we obtain the number of degrees of freedom
inside the sphere
\begin{equation}
N_{bulk}=\frac{-16\pi^{2}(\rho+3p)}{3H^{4}}.
\end{equation}
Using the relation Eq.(3), we know the number of degrees of freedom
on the surface:
\begin{equation}
N_{sur}=\frac{A}{L_{P}^{2}}+\frac{4\pi}{\omega}\ln\frac{A}{4L_{P}^{2}}.
\end{equation}
Then we can obtain the dynamic equation by using the Eq.(9)
\begin{equation}
\dot{H}+H^{2}=-\frac{L_{p}^{2}H^{4}}{\omega}\ln\frac{\pi}{H^{2}L_{p}^{2}}-\frac{4\pi L_{p}^{2}}{3}(\rho+3p).
\end{equation}
Since $\frac{\ddot{a}}{a}=\dot{H}+H^{2}$, we have the equation:
\begin{equation}
\frac{\ddot{a}}{a}=-\frac{L_{p}^{2}H^{4}}{\omega}\ln\frac{\pi}{H^{2}L_{p}^{2}}-\frac{4\pi L_{p}^{2}}{3}(\rho+3p).
\end{equation}
From this equation, we can know the standard dynamic equation of FRW
universe in the GR gravity will be recovered when the parameter $\omega\rightarrow\infty$.
Here if we use Eq.(2) in place of Eq.(3), we can't obtain the correct
result, because degrees of freedom on the surface don't uniformly
distribute over the area. As the $f(R)$ theory, the total energy-momentum
tensor $T_{\mu\nu}$ can't be determined if the form
of deformed HL gravity is not given, so we will use the form of deformed HL gravity
to determine it. Further, the form of entropy Eq.(29) is also determined
by the explicit form of deformed HL gravity.

Now we recall the HL theory, this gravity has the ADM formalism\cite{key-18,key-19}
\begin{equation}
ds^{2}=-N^{2}dt^{2}+g_{ij}(dx^{i}+N^{i}dt)(dx^{j}+N^{j}dt).
\end{equation}
The action of HL gravity can be written
\begin{equation}
S_{HL}=\int dtd^{3}xN\sqrt{g}(L_{K}+L_{V}+L_{M}).
\end{equation}
\begin{equation}
L_{K}=\frac{2}{\kappa^{2}}(K_{ij}K^{ij}-\lambda K^{2}),
\end{equation}
\begin{equation}
L_{V}=\frac{\kappa^{2}\mu^{2}}{8(1-3\lambda)}\left(\frac{1-4\lambda}{4}R^{2}+\Lambda_{W}R-3\Lambda_{W}^{2}\right)-\frac{\kappa^{2}}{2w^{4}}\left(C_{ij}-\frac{\mu w^{2}}{2}R_{ij}\right)\left(C^{ij}-\frac{\mu w^{2}}{2}R^{ij}\right),
\end{equation}
where $L_{K}$ is the kinetic term, $L_{V}$ is the potential term,
and $L_{M}$ is the matter term. The extrinsic curvature and Cotten
tensor are given by
\begin{equation}
K_{ij}=\frac{1}{2N}(\dot{g}-\nabla_{i}N_{j}-\nabla_{j}N_{i}),
\end{equation}
\begin{equation}
C^{ij}=\epsilon^{ikl}\nabla_{k}\left(R_{l}^{j}-\frac{1}{4}R\delta_{l}^{j}\right).
\end{equation}
In the infrared(IR) limit, the action should be reduced to the EH
action of the general relativity
\begin{equation}
S_{EH}=\frac{1}{16\pi G}\int d^{4}xN\sqrt{g}(K_{ij}K^{ij}-K^{2}+R-2\Lambda)
\end{equation}
by setting $x^{0}=ct$, $\lambda=1$. The speed of light, the Newton's
constant and the cosmological constant are respectively given by
\begin{equation}
c=\frac{\kappa^{2}\mu}{4}\sqrt{\frac{\Lambda_{W}}{1-3\lambda},}G=\frac{\kappa^{2}}{32\pi c},\Lambda=\frac{3}{2}\Lambda_{W}.
\end{equation}
By introducing a soft violation term $\mu^{4}R$, it is called as
the ``deformed HL gravity''~\cite{key-18}. The speed of light and
Newton's constant in the IR limit are given by
\begin{equation}
c^{2}=\frac{\kappa^{2}\mu^{4}}{2},G=\frac{\kappa^{2}}{32\pi c},\lambda=1.
\end{equation}
Based on these, the entropy Eq.(29) can be obtain\cite{key-18}.

As in the paper\cite{key-19,key-20}, we define the energy and pressure
of this universe
\begin{equation}
\rho\equiv\rho_{m}+\rho_{\Lambda}+\rho_{k}+\rho_{dr},p\equiv p_{m}+p_{\Lambda}+p_{k}+p_{dr}.
\end{equation}
Where we use the energy $\rho_{m}$ and pressure $p_{m}$ of perfect
fluid as the ones of the matter. By introducing the cosmological constant
term, the curvature term and the dark radiation term\cite{key-19}
\begin{equation}
\rho_{\Lambda}=-p_{\Lambda}=-\frac{3\kappa^{2}\mu^{2}\Lambda_{W}^{2}}{8(3\lambda-1)}
\end{equation}
\begin{equation}
\rho_{k}=-3p_{k}=\frac{3k}{4(3\lambda-1)a^{2}}\left(\kappa^{2}\mu^{2}\Lambda_{W}-8\mu^{4}(3\lambda-1)+\frac{8}{\kappa^{2}}(3\lambda-1)^{2}\right)
\end{equation}
\begin{equation}
\rho_{dr}=3p_{dr}=\frac{3\kappa^{2}\mu^{2}}{8(3\lambda-1)}\frac{k^{2}}{a^{4}}
\end{equation}
So far, we have already obtained the complete dynamic equation of
FRW universe by Eq.(33) together with (43). The dynamic equation is
derived from the idea of emergence of space. The deformed HL gravity
is a quantum gravity theory, hence it describes self-consistently
the physical results which contains the quantum effects. Then the
Eq.(33) should be seen as the effective dynamic equation of FRW universe
which contain the effective quantum gravitational effects.

The above results show that if
we consider the evolution of universe in the general gravitational
theories, the evolution of universe can be described by Eq.(9) according to the Padmanabhan's
idea. The choice of $N_{sur}$ is important in the Eq.(9). In general,
we can use the form of the entropy of the black hole to describe that
of the spacetime due to their similarity near the horizon. We give
the relation between the number of the surface degrees of freedom
and the entropy of the horizon of the spacetime , and take the specific
forms of the $f(R)$ gravity and deformes HL gravity to determine
their effective energy-momentum tensors and the entropy. Furthermore,
we can also obtain the corresponding dynamical equation of FRW universe
by the equation of evolution Eq.(9).

\section*{4. Discussion and conclusion}

In this paper, our idea mainly comes from\cite{key-9}the thought
that the expansion of universe is due to the difference of the number
of degrees of freedom on the spherical surface and the one inside
the sphere. We use also $H^{-1}$as the radius of the expanding universe,
and take $(\rho+3p)<0$ in order to have the asymptotic holographic
equipartition. This also implies the existence of the dark energy.
We obtain the dynamic equation of FRW universe from the idea of emergence
of spaces in the $f(R)$ theory and deformed HL theory, and confirm
Padmanabhan's idea by these examples. It should be emphasized that
it is the relation $N_{sur}=\frac{4S_{H}}{k_{B}}$ that describes
the number of degrees of freedom on the surface of the horizon, because
the number of degrees of freedom are described by the entropy in the
statistical physics. Hence this relation is natural and general .

By establishing the relationship between the number of degrees of
freedom on the spherical surface and the entropy of the black hole,
using the equipartition law and the evolution equation(9), we have
indeed obtained the dynamic equation of FRW universe in the $f(R)$
theory and deformed HL theory. Here the Komar energy plays an important
role which is taken as the usual form for gravitational acceleration
$|E|=2\left|\int_{\Sigma}\left(T_{\mu\nu}-\frac{1}{2}Tg_{\mu\nu}\right)U^{\mu}U^{\nu}dV\right|$.
The reason why we can\textquoteright{}t simply take the energy-momentum
tensor of the perfect fluid as the total energy-momentum tensor is
the form of the entropy of the horizon of the spacetime and the total
energy-momentum tensor are different in the different theories. Hence
one has to use the specific form of gravitational theory in order
to obtain the corresponding dynamical equation of FRW universe. Then,
we would also like to explain Eq.(1) again. Because of $S=\frac{k_{B}A}{4L_{P}^{2}}$
in the GR theory and some gravity theories, Eq.(1) can be transformed to Eq.(2)
in those theories. But in some theories such as the $f(R)$
theory and deformed HL theory with $S\neq\frac{k_{B}A}{4L_{P}^{2}}$,
one can't use Eq.(2), because degrees of freedom on the surface don't
uniformly distribute over the area. Therefore our relation(Eq.(1))
is more general than Eq.(2), especially in quantum gravity theory it can be widely applied.
Our results are useful for further understanding of the expansion
of the universe in general gravity theories.


\begin{thebibliography}{10}

\bibitem{key-1}J.M. Bardeen, B. Carter and S.W. Hawking, Comm, Math.
Phys. 31 161 (1973).

\bibitem{key-2}S.W. Hawking, Comm. Math. Phys. 43 199 (1975).

\bibitem{key-3}T. Jacobson, Phys.Rev. Lett. 75, 1260 (1995).

\bibitem{key-4}T. Padmanabhan, Class.Quant.Grav.19:5387-5408(2002)

\bibitem{key-5}E. P. Verlinde, JHEP 1104, 029 (2011).

\bibitem{key-6}R. -G. Cai and S. P. Kim, JHEP 02 (2005) 050.

\bibitem{key-7}M. Akbar and R. -G. Cai, Phys. Lett. B 635(2006)7-10.

\bibitem{key-8}S. -W. Wei, Y. -X. Liu, Y. -Q. Wang, Commun.Theor.Phys.56:455-458(2011)

\bibitem{key-9}T. Padmanabhan, arXiv:1206.4916.

\bibitem{key-10}R. -G. Cai, JHEP11(2012)016.

\bibitem{key-11}T. Padmanabhan, Research in Astron. Astrophys 12,
8, 891\textendash{}916(2012)

\bibitem{key-12}T. Padmanabhan, AIP Conf. Proc. 1483, 212-238 (2012)

\bibitem{key-13}T. Padmanabhan, Rep. Prog. Phys. 73 (2010) 046901.

\bibitem{key-14}T. Padmanabhan, Class.Quant.Grav.21:4485-4494(2004).

\bibitem{key-15}R. M. Wald, Phys.Rev. D 48, 3427(1993).

\bibitem{key-16}S. Weinberg, Cosmology, Oxford university press(2008).

\bibitem{key-17}S. Capozziello, Int. J. Mod. Phys. D 11, 483(2002);
S. Capozziello,V. F. Cardone, S. Carloni, Int. J. Mod. Phys. D 12,
1969(2003).

\bibitem{key-18}Y. S. Myung, Phys. Lett. B 684(2010) 158-161.

\bibitem{key-19}Q. -J. Cao, Y. -X Chen, K. -N. Shao, JCAP05(2010)030.

\bibitem{key-20}A. -Z. Wang and Y. -M Wu, JCAP07(2009)012.

\end{thebibliography}
\end{document}